\documentclass[conference,a4paper]{IEEEtran}
\usepackage{xcolor}
\usepackage{balance}

\usepackage{amsmath}
\usepackage{float}
\usepackage[style=ieee]{biblatex}
\ifCLASSINFOpdf
  \usepackage[pdftex]{graphicx}
\else
  \usepackage[dvips]{graphicx}
\fi
\graphicspath{{img/}}

\ifCLASSOPTIONcompsoc
 \usepackage[caption=false,font=normalsize,labelfont=sf,textfont=sf]{subfig}
\else
 \usepackage[caption=false,font=footnotesize]{subfig}
\fi
\begin{document}
%
\title{On Models and Approaches for Human Vital Signs Extraction from Short Range Radar Signals}

\author{\IEEEauthorblockN{
Mikolaj Czerkawski,
Christos Ilioudis,
Carmine Clemente,
Craig Michie,
Ivan Andonovic,
Christos Tachtatzis
}
                                    % ...

\IEEEauthorblockA{
Department of Electronic and Electrical Engineering, University of Strathclyde, Glasgow G1 1XW, UK}
}



\maketitle

\begin{abstract}
The paper centres on an assessment of the modelling approaches for the processing of signals in CW and FMCW radar-based systems for the detection of vital signs. It is shown that the use of the widely adopted phase extraction method, which relies on the approximation of the target as a single point scatterer, has limitations in respect of the simultaneous estimation of both respiratory and heart rates. A method based on a velocity spectrum is proposed as an alternative with the ability to treat a wider range of application scenarios.
\end{abstract}

\vskip0.5\baselineskip
\begin{IEEEkeywords}
 Doppler processing, FMCW radar, heart rate monitoring, micro-Doppler, millimeter wave radar, non-contact monitoring, respiration rate monitoring, vital signs.
\end{IEEEkeywords}

%

\section{Introduction}
    The remote detection of vital signs from Doppler derived measurements using radar has been subject to significant research~\cite{1975uwaverespiration, 2002dsp_vs, 2008novelsp, 2013advdopreview}. The capability to detect both respiration and heart rate without physical contact represents the foundation for the development of a number of applications~\cite{2009arrythmia, 2013advdopreview, 2018breathingdisorder, 2018sleepstage}, ranging from monitoring patients with specific conditions, such as sleep apnea, to patients where constant monitoring of the vital signs is advised due to risks of the onset of life-threatening conditions.
    
    Early implementations of non-contact vital sign monitoring were based on CW radar technology. Recently more compact, low-cost Frequency Modulated Continuous Wave (FMCW) platforms with the capability to map the environment in range and velocity have become more readily available. Regardless of radar  mode, the approach to the processing of signals has remained relatively unchanged.
    
    Reported FMCW-based vital sign detection solutions to date, have relied on a traditional phase extraction approach~\cite{2015fmcw80,2016bodymov, 2019multiplepatients, 2019remote77}. The phase of the reflection vector is extracted, unwrapped and interpreted as changes in range, proportional to the mechanical motions induced by in this application, organ contractions. At its core is the assumption that the target may be faithfully represented as a single point scatterer, demonstrated to be inappropriate when both of these vital signs are of interest.
    
    The paper investigates the effectiveness of the phase-based approach to estimate vital signs; the interpretation of the signal is significantly more challenging if the model does not represent the observed target sufficiently well. Two approaches to modelling vital signs using a (i) single point scatterer and (ii) two point scatterers models are assessed, described in Section~\ref{sec:modelling}. Section~\ref{sec:sim_results} presents a simulation-based evaluation of the effectiveness of the extraction of the phase for the two models in the goal of deriving both respiratory and heart rate simultaneously. Key conclusions are drawn within Section~\ref{sec:conclusion}.

\section{Modelling Approaches}
\label{sec:modelling}
    Breathing and heart rates may be described as two dynamic components $R_b(t)$ and $R_h(t)$, equal to the local changes in distance owing to each motion. Both motions are normally periodic but with different amplitudes, harmonic content (affecting the shape of the motion trace) and frequencies.
    
    The simplest approach to modelling vital signs is to assume an offset-radius $R_0$ corresponding to the distance to the centre of the target and to add to it both of the dynamic components:
    \begin{align}
        R(t) = R_0 + R_b(t) + R_h(t)
        \label{eq:distance}
    \end{align}
    The inherent assumption is that the animate object can be modelled by a single point scatterer, proven to be an adequate approximation for scenarios where both vital signs are present with a preserved ratio of amplitudes $\frac{A_b}{A_h}$ in each section of the target surface. However, this is rarely the case. Lung inflation stretches a wide portion of the torso, resulting in high amplitude motions ($A_b$), whilst the heartbeat motion is predominately propagated by the cardiovascular system. Consequently, under these conditions, only the respiratory motion is readily detectable.
    
    The limitation has motivated research to estimate the Radar Cross-Section (RCS) for each of the human vital signs~\cite{2010rcs, 2011rcs}, with the conclusions that the RCS for respiration is consistently and significantly larger than that for heartbeat. Furthermore,~\cite{2017vsdistribution} report that the manifestation of each vital sign is distributed differently over the body, further evidence that the modelling of the target as a single scattering point has limitations.
    
    The model for a baseband received signal can be written as:
    \begin{align}
        \label{eq:s_sp}
        s(t) = |X_0| e^{j\frac{4\pi}{\lambda}(R_b(t) + R_h(t))}
    \end{align}
    where $|X_0|$ is the effective received spectrum magnitude due to the superposition of a number of scattering points\footnote{At this point, $R_0$ contribution to the signal may be omitted in all equations considered in this analysis since it only results in a constant phase offset due to multiplying the received signal by $e^{j\frac{4\pi}{\lambda}R_0}$. Ultimately, only the dynamic components of the range trace are of interest.} and $\lambda$ represents the carrier wavelength. Equation \eqref{eq:s_sp} may lead to a limited description of what is being observed, especially in cases with a significant difference in RCS between vital signs.
    
    Therefore, a more appropriate approach is to model the vital signs as two distinct point scatterers:
    \begin{align}
         \label{eq:s_tp}
        s(t) = |X_r|e^{j\frac{4\pi R_b(t)}{\lambda}} + |X_h|e^{j\frac{4\pi R_h(t)}{\lambda}}
    \end{align}
    where $|X_r|$ and $|X_h|$ represent the effective magnitudes corresponding to each vital sign.
    
    Previous research executed phase extraction by applying an inverse tangent function to the quadrature components followed by the unwrapping of the obtained vector~\cite{2015fmcw80,2016bodymov, 2019multiplepatients, 2019remote77}, approximating the target as a single point scatterer described by ~\eqref{eq:s_sp}. The dynamic phase,~$\phi_{1p}(t)$ reduces to the sum of the two vital sign contributions:
    \begin{align}
        \label{eq:ph_sp}
        &\phi_{1p}(t)=\frac{4\pi}{\lambda}\left[R_b(t)+R_h(t)\right]
    \end{align}
    The two point model, on the other hand, leads to a substantially different dynamic phase term, $\phi_{2p}(t)$:
    \begin{align}
        \label{eq:ph_tp}
        &\phi_{2p}(t)=\nonumber\\
        &\textrm{arctan}\left(\frac{|X_r|\sin(\frac{4\pi}{\lambda} R_b(t))+|X_h|\sin(\frac{4\pi}{\lambda} R_h(t))}{|X_r|\cos(\frac{4\pi}{\lambda} R_b(t))+|X_h|\cos(\frac{4\pi}{\lambda} R_h(t))}\right)
    \end{align}
    It is clear that for a point scatterer model, it is possible to fully recover all of the dynamic range information. However, for a case where the two vital signs are not distributed equally over the surface, the phase term is no longer a linear combination of $R_b(t)$ and $R_h(t)$.
    
    The respiratory motion is not only of much higher amplitude but is also present over a wider area than the heartbeat-induced motions, leading to $|X_r| >> |X_h|$. In that case, the phase term for the two point scatterers model approaches the limit of $\frac{4\pi}{\lambda}R_b(t)$. At the same time, the heartbeat component in the signal extracted from the phase becomes negligible.
    
    Equation ~\eqref{eq:s_tp} describes the dynamic content of $s(t)$ as a sum of complex exponential components. Fourier Transformation (FT) then becomes a suitable decomposition method. Using Short-Time FT and converting the Doppler frequency to velocity, generates a matrix of velocity spectra distributed over time, referred to as velocity-time maps.
    
    A simulation environment that generates synthetic motion signals corresponding to each vital sign was established and these signals were subsequently analysed to assess the performance of the two processing methods under consideration.
    
\section{Simulation Results}
\label{sec:sim_results}
\begin{figure}[!thb]
        \centering
        \includegraphics[width=\linewidth]{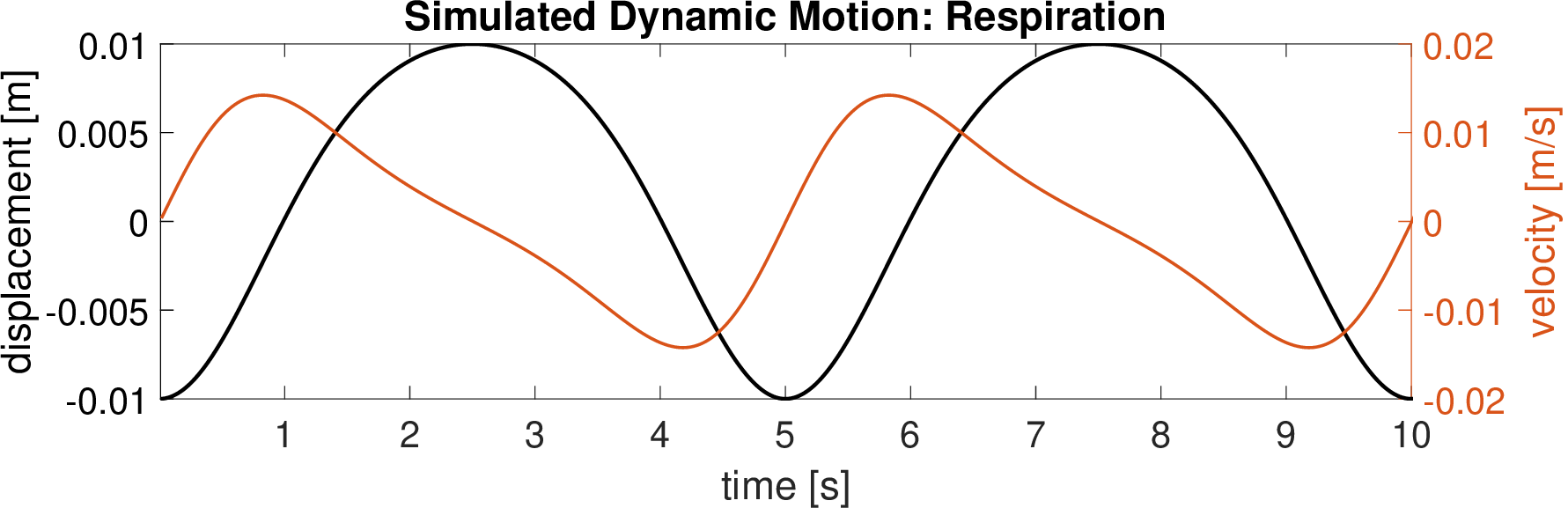}
        \caption{Simulated Displacement due to Respiration}
        \label{fig:resp_disp}
    \end{figure}

\begin{figure}[!thb]
        \centering
        \includegraphics[width=\linewidth]{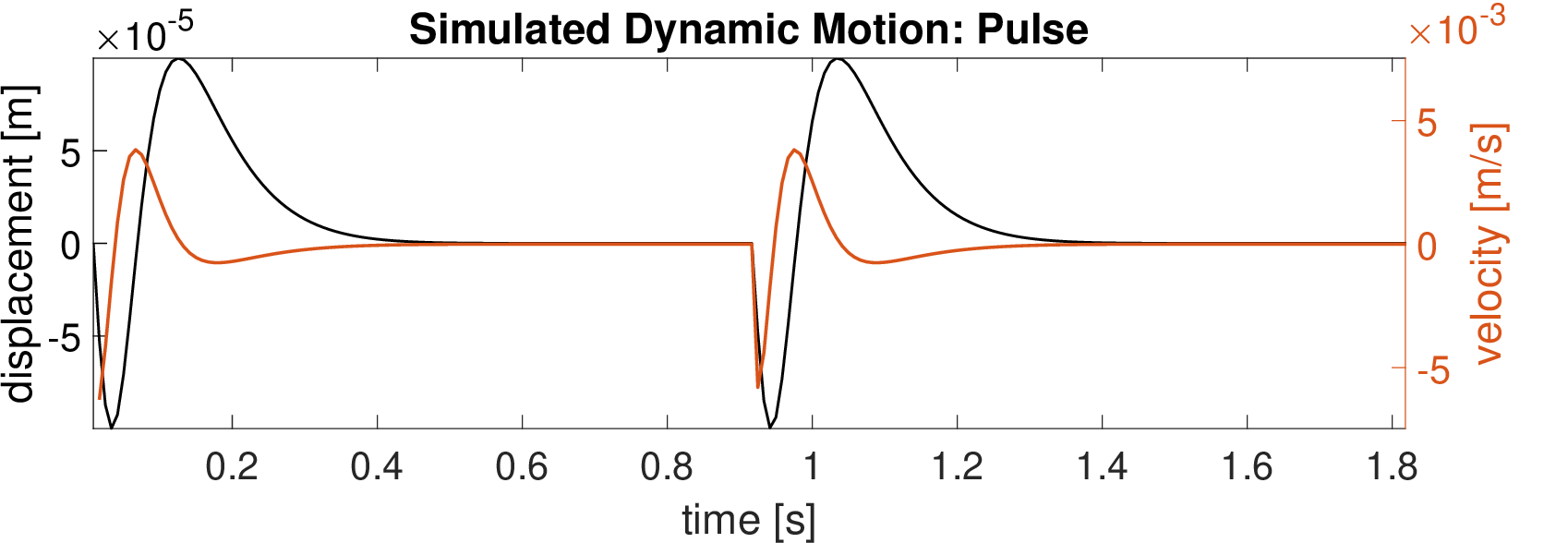}
        \caption{Simulated Displacement due to Heart Activity}
        \label{fig:heart_disp}
    \end{figure}

    The simulation was carried out for constant vital sign rates and a V-Band Radar carrier of 60~GHz, with the settings listed in Table~\ref{tab:simulation_settings}. The rates of the vital signs were selected such that none of the harmonics of the respiratory component interfere with the heart rate spectrum.
    \begin{table}[tb]
        \centering
        \caption{Simulated System Settings}
        \begin{tabular}{l l}
            \hline
            Respiration Rate, $f_b$ & 0.2 Hz \\
            Heart Rate, $f_h$ & 1.1 Hz \\
            H-R Ratio & -10 dB\\
            Sample Rate & 120 sps\\
            Carrier Frequency & 60 GHz \\
            Respiratory Motion Amplitude, $A_b$ & 1.0 cm\\
            Heartbeat Motion Amplitude, $A_h$ & 0.1 mm\\
            \hline
        \end{tabular}
        \label{tab:simulation_settings}
    \end{table}
    The synthesis of the signal utilises two equations in order to obtain curves representative of the real vital signs. The applied approximations for the range displacements have been informed by a significant number of experiments capturing motions owing to the heartbeat \cite{1987mechanocardiogram}, or  respiration \cite{2008capresp}.
    \begin{align}
        \label{eq:sim_resp}
        &R_b(t) = A_b\frac{5-2^{2+\cos(2 \pi \int_{-\infty}^tf_b(t) dt)}}{3}\\
        \label{eq:sim_pulse}
        &R_h(t) = A_h \sin\left(2\pi \text{mod}(-\int_{-\infty}^tf_h(t)dt, 1)^{10}\right)
    \end{align}
    Each equation can be expressed as a product of an amplitude $A_b$, $A_h$ and a term that oscillates between -1 and 1, dependent on the corresponding vital sign rate $f_b$ or $f_h$. The purpose of~\eqref{eq:sim_resp} and~\eqref{eq:sim_pulse} is to emulate the harmonic content of each vital sign signal, observed in the time domain as a bell-like shape of the respiration curve (\figurename~\ref{fig:resp_disp}), or the impulse property of the heart rate signal (\figurename~\ref{fig:heart_disp}).
    \\
    
\begin{figure}[!tb]
        \centering
        \includegraphics[width=\linewidth]{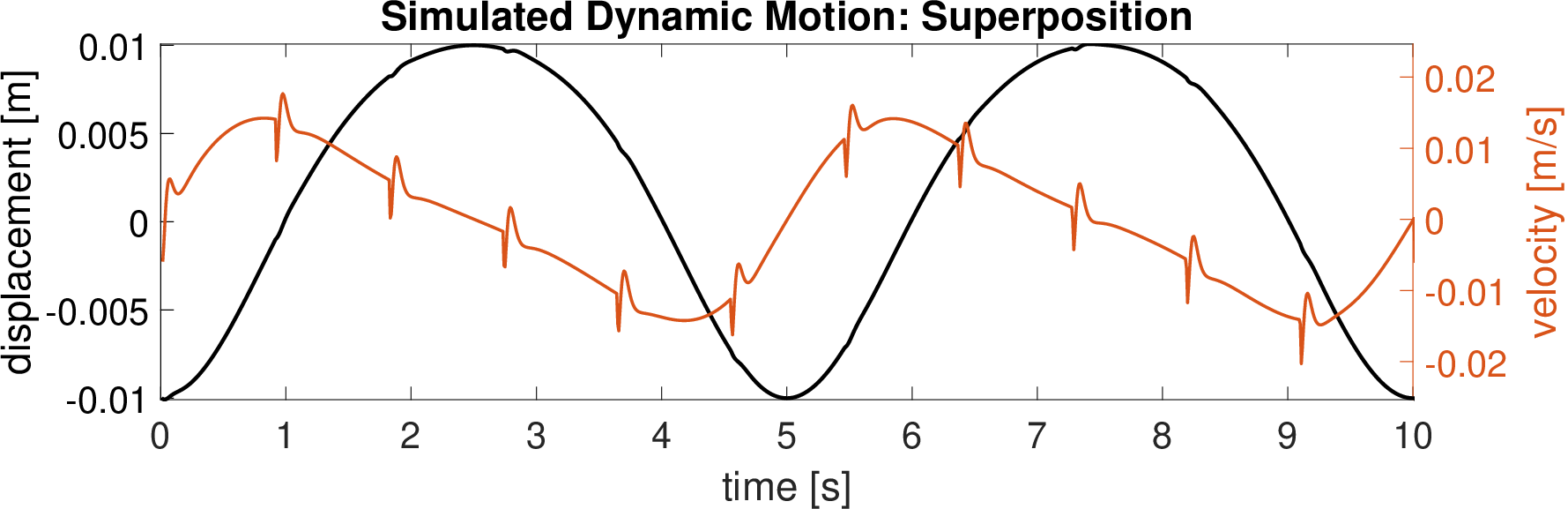}
        \caption{Sum of the Simulated Displacements}
        \label{fig:total_disp}
    \end{figure}

    The single point scatterer model assumes that these two motions contribute to the reflected signal phase equally, as if both motions were superimposed across the entire target (\figurename~\ref{fig:total_disp}) and reflected.
    
    An analysis of the synthesised signals was carried out to determine their impact on the Doppler spectrum. The signals are described by~\eqref{eq:sim_resp} and~\eqref{eq:sim_pulse} and were substituted as $d(t)$ to the phase of a complex reflection vector $s(t) = e^{j\frac{4\pi}{\lambda}d(t)}$.
    
\begin{figure}[!tb]
        \centering
        \includegraphics[width=\columnwidth]{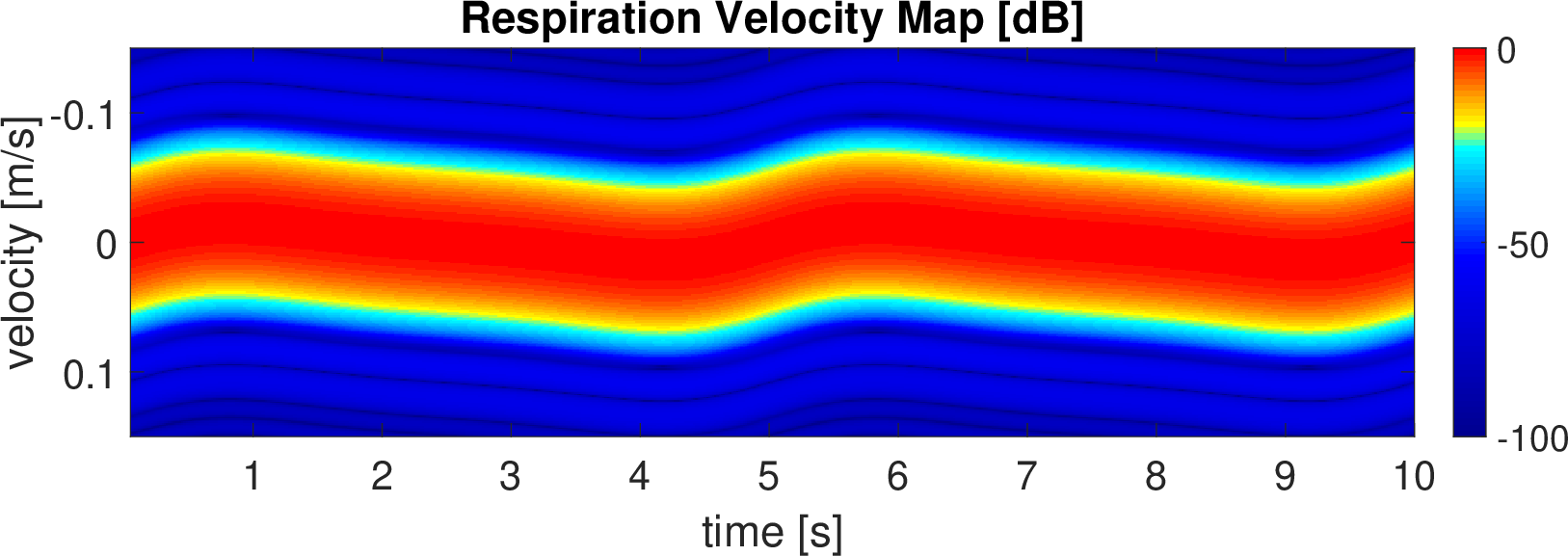}
        \caption{Velocity Time Map of the synthesised Respiration Component}
        \label{fig:resp_dopp}
    \end{figure}

\begin{figure}[!t]
        \centering
        \includegraphics[width=\columnwidth]{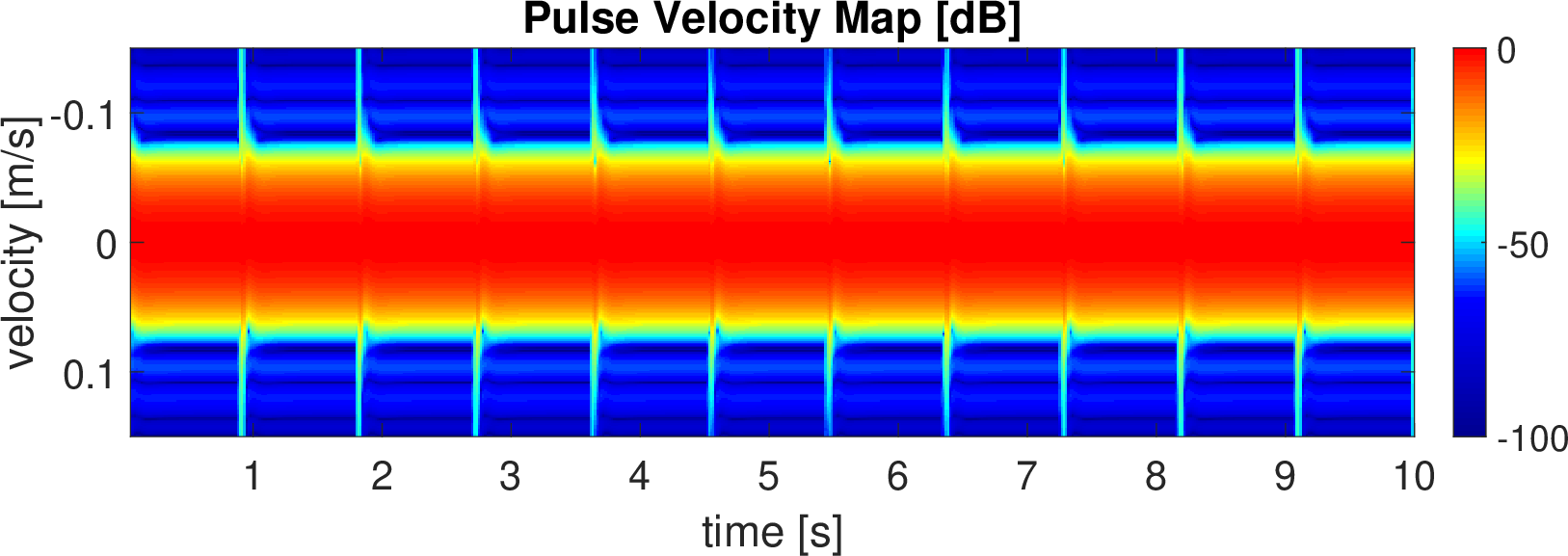}
        \caption{Velocity Time Map of the synthesised Pulse Component}
        \label{fig:pulse_dopp}
    \end{figure}

    The analysis of the simulated data was executed at a 0.1~second Doppler-analysis window applied over a 60~seconds recording. \figurename~\ref{fig:resp_dopp} and \figurename~\ref{fig:pulse_dopp} show the resultant velocity-time maps of both synthesised vital signs. The velocity translation is carried out using the Doppler equation $v = -\frac{\lambda}{2}f_D$. The simulated respiration signal is relatively strong and evident is that the shape contained in \figurename~\ref{fig:resp_dopp} matches the shape of the derivative curve (velocity) of the respiratory motion signal. The heartbeat velocity-time map is harder to interpret because of the low amplitude generated by motion of the heart. Even though the motion is weak for the whole derived shape to be clearly visible, a velocity impulse corresponding to the original signal is nevertheless apparent every 1.1~seconds.
    
    For the simultaneous detection of the vital signs, the two modelling frameworks along with two processing methods were compared to inform the selection of the optimum approach for the application under development.
    
\begin{figure}[!t]
        \centering
        \includegraphics[width=\columnwidth]{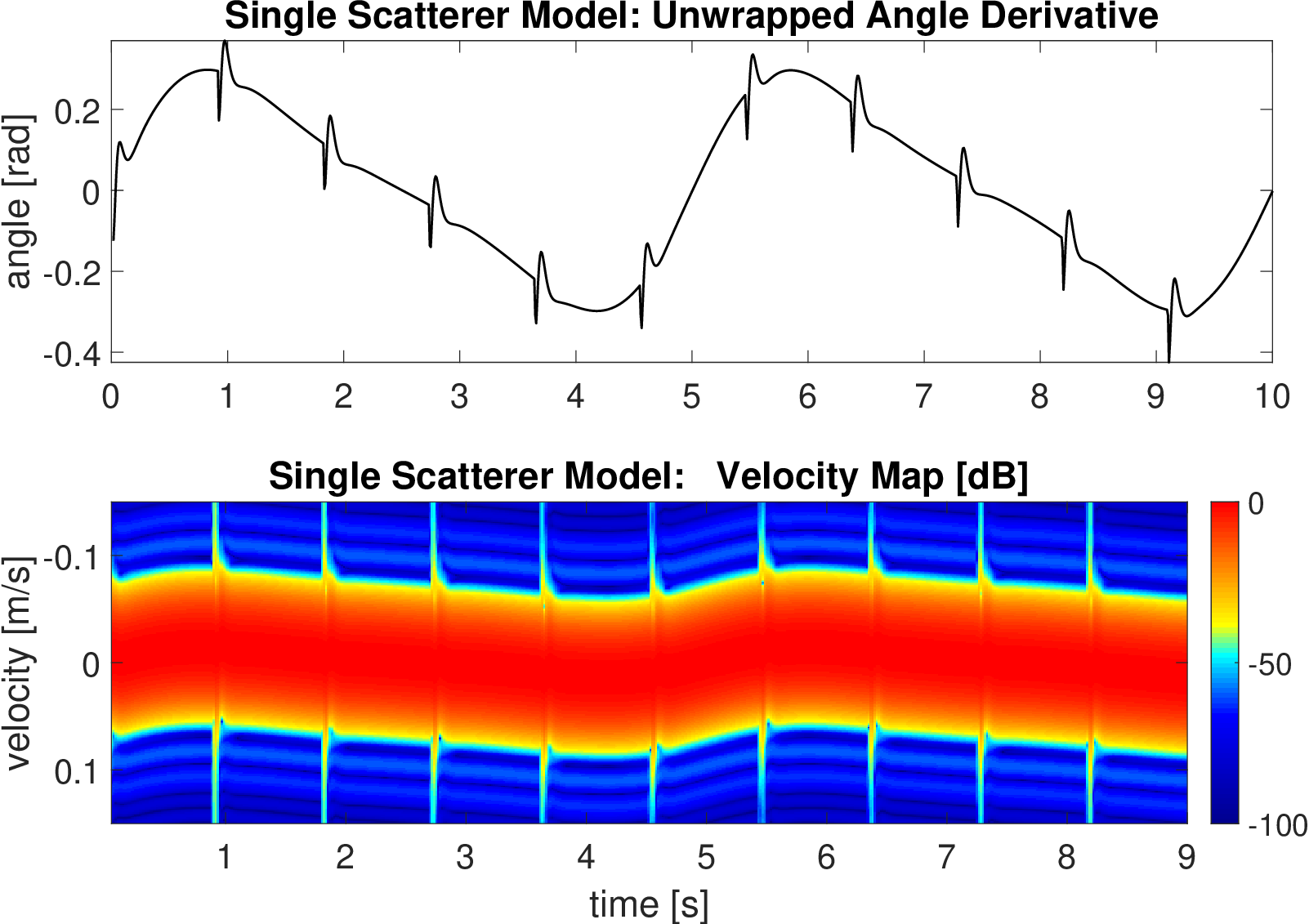}
        \caption{Processing Output of Single Point Scatterer Signal}
        \label{fig:single_scatterer}
    \end{figure}

    \figurename~\ref{fig:single_scatterer} shows the result based on a single point scatterer moving according to both vital sign motions simultaneously. The motion can be faithfully reconstructed by extracting the phase of the complex signal and unwrapping. The top plot shows the unwrapped phase, proportional to the detected velocity. The alternative is to compute the velocity-time map, as shown in the bottom graph. It is clear that rates of both vital signs can be extracted from the map, the respiration as the low-frequency modulation of low velocities whilst the heart rate as the higher frequency impulse modulation of high velocities. Both processing techniques can estimate the rates with the phase extraction technique from a single point scatterer being less computationally expensive than the velocity-time map analysis.
    
\begin{figure}[!b]
        \centering
        \includegraphics[width=\columnwidth]{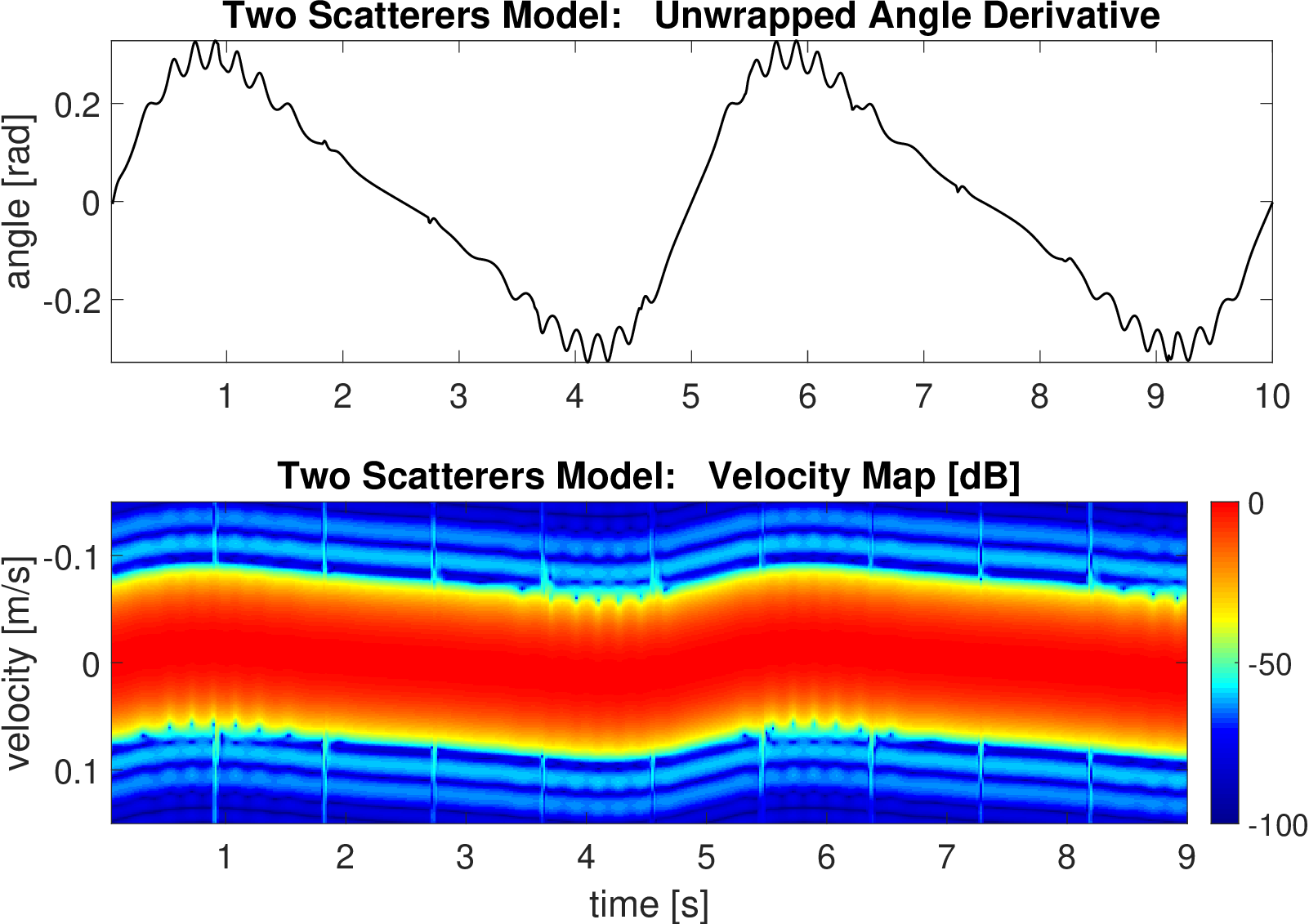}
        \caption{Processing Output of Two Point Scatterers Signal}
        \label{fig:two_scatterers}
    \end{figure}

    The challenge of extracting vital sign rates becomes more difficult when the target resembles two separate point scatterers. Since the heart rate signal is not only of lower amplitude but also present over a much smaller area, its contribution to the net phase of the complex signal may be negligible. As is evident in \figurename~\ref{fig:two_scatterers}, the extracted phase derivative is distorted and whilst the respiratory motion velocity is preserved, the heart rate component is less apparent.
    
\begin{figure}[!t]
        \centering
        \includegraphics[width=\columnwidth]{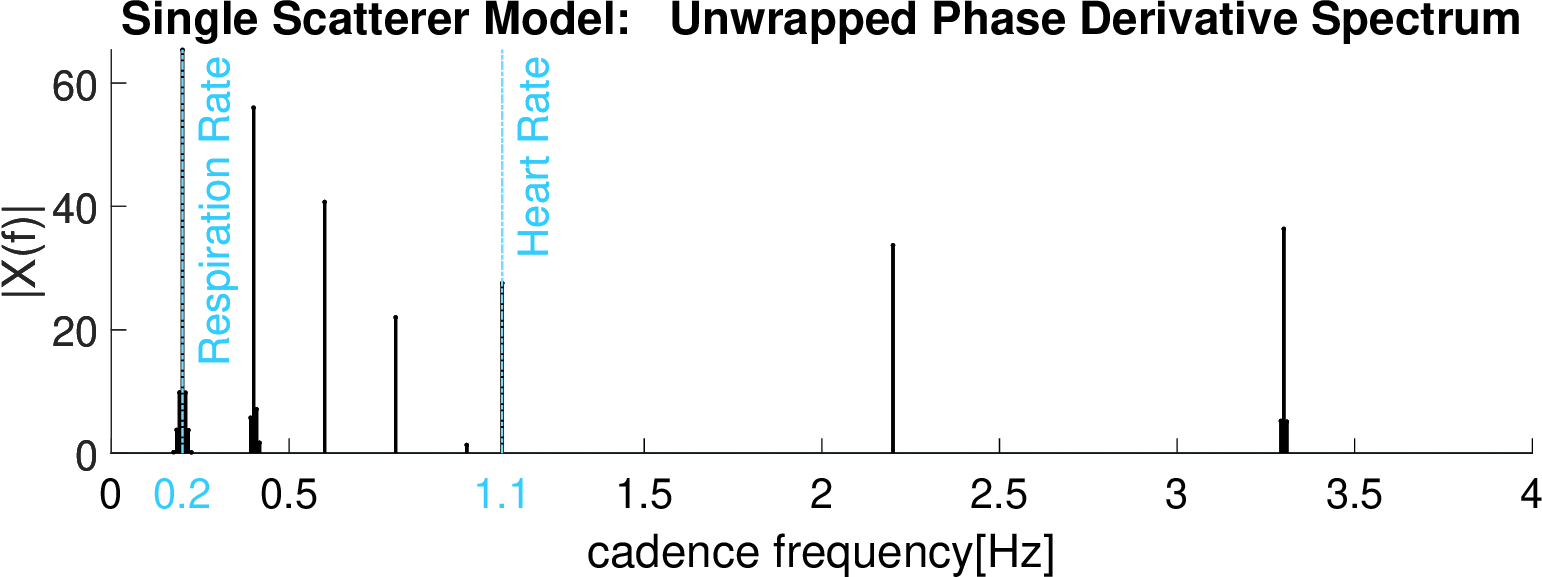}
        \caption{Spectrum of the Unwrapped Phase of the Single Point Scatterer}
        \label{fig:single_scatterer_phase_spec}
    \end{figure}

\begin{figure}[!t]
        \centering
        \includegraphics[width=\columnwidth]{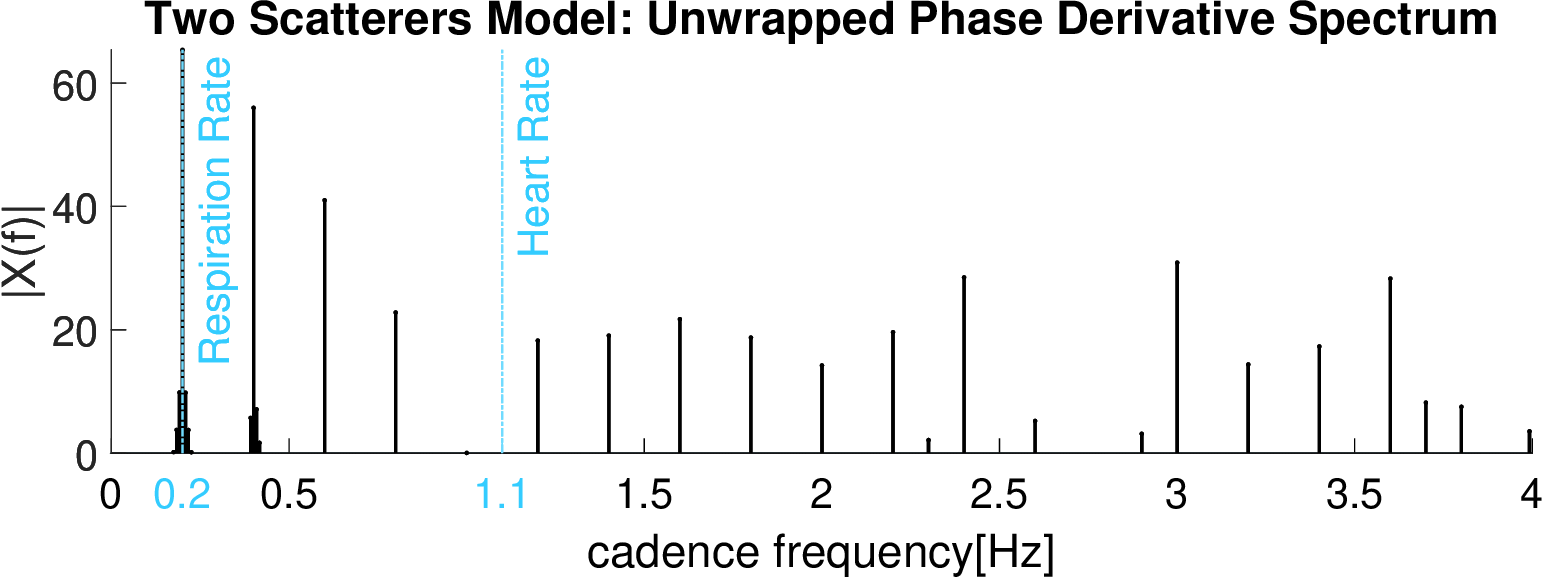}
        \caption{Spectrum of the Unwrapped Phase of Two Point Scatterers}
        \label{fig:two_scatterers_phase_spec}
    \end{figure}

    \figurename~\ref{fig:single_scatterer_phase_spec} and~\ref{fig:two_scatterers_phase_spec} show the spectrum of the unwrapped phase for each of the two models. For the two point scatterers model, the phase spectrum can only be used to extract the respiration rate. In the case of a single point scatterer, the results are clear, comprising a series of respiratory components at every 0.2~Hz, and a series of heart rate harmonics at every 1.1~Hz. By arranging the strong components into harmonic groups, both vital signs can be reliably extracted.
    
    The limitation of the phase extraction technique can be contrasted with results using velocity-time map (bottom graph of \figurename~\ref{fig:two_scatterers}). Although the impulse component in the high-velocity band is not as strong as in the case of the single point scatterer model (since the heart complex reflection component is 10~dB lower than the respiration vector), it is nevertheless possible to observe impulses at the period of the heartbeat.
    
\begin{figure}[!b]
        \centering
        \includegraphics[width=\columnwidth]{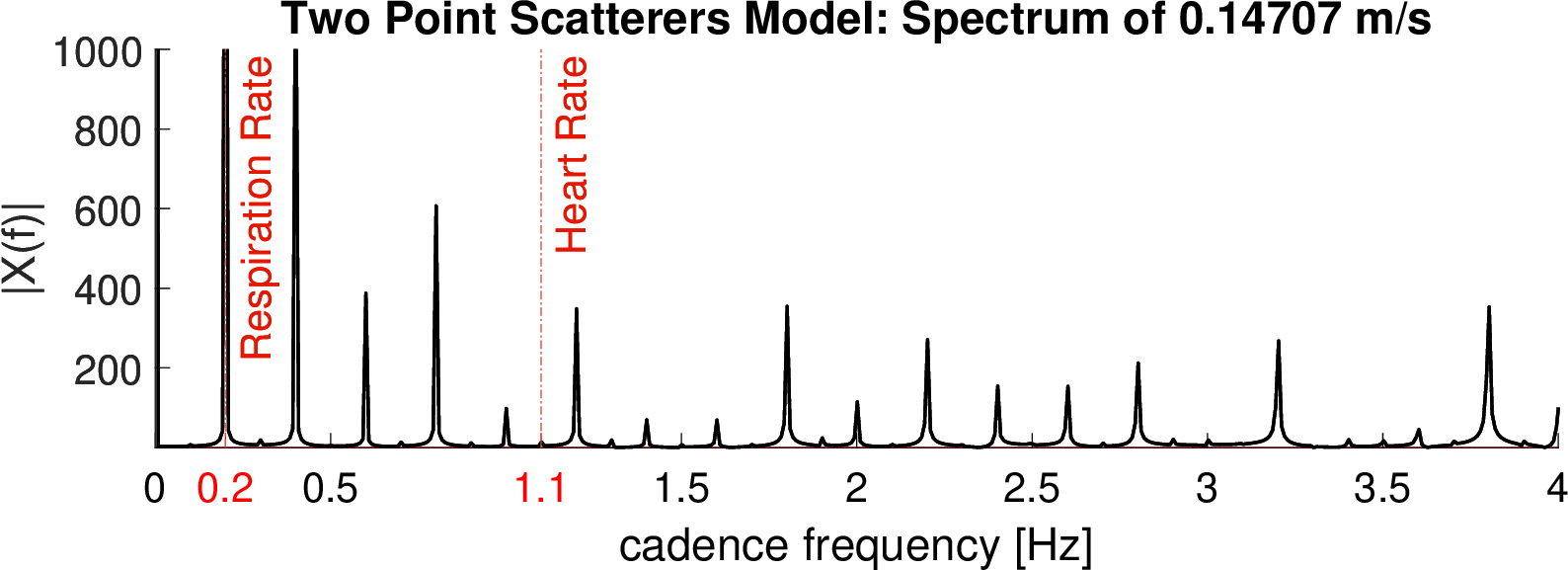}
        \caption{Spectrum of 14.7 cm/s for the Two Point Scatterers}
        \label{fig:two_scatterers_highvel_alt}
    \end{figure}

    Further, the high bandwidth impulses due to the heartbeat that are most apparent for high velocities can be used to extract a time-domain spectrum magnitude vector for a selected velocity. For illustration, 14.7~cm/s was chosen. Applying~an~FT to this time signal leads to the result presented in \figurename~\ref{fig:two_scatterers_highvel_alt}. With this approach, the heartbeat component of the signal is dominated by respiratory harmonics. One solution is to apply a logarithm to the vector in the time domain before computing the spectrum, an operation equivalent to computing the spectrum of the values shown in \figurename~\ref{fig:two_scatterers} (bottom) since all of the spectrograms in the paper are plotted in dB scale. The logarithm operation is capable of compressing the dynamic range and emphasising the impulsive components of the signal.
    
    \begin{figure}[!t]
        \centering
        \includegraphics[width=\columnwidth]{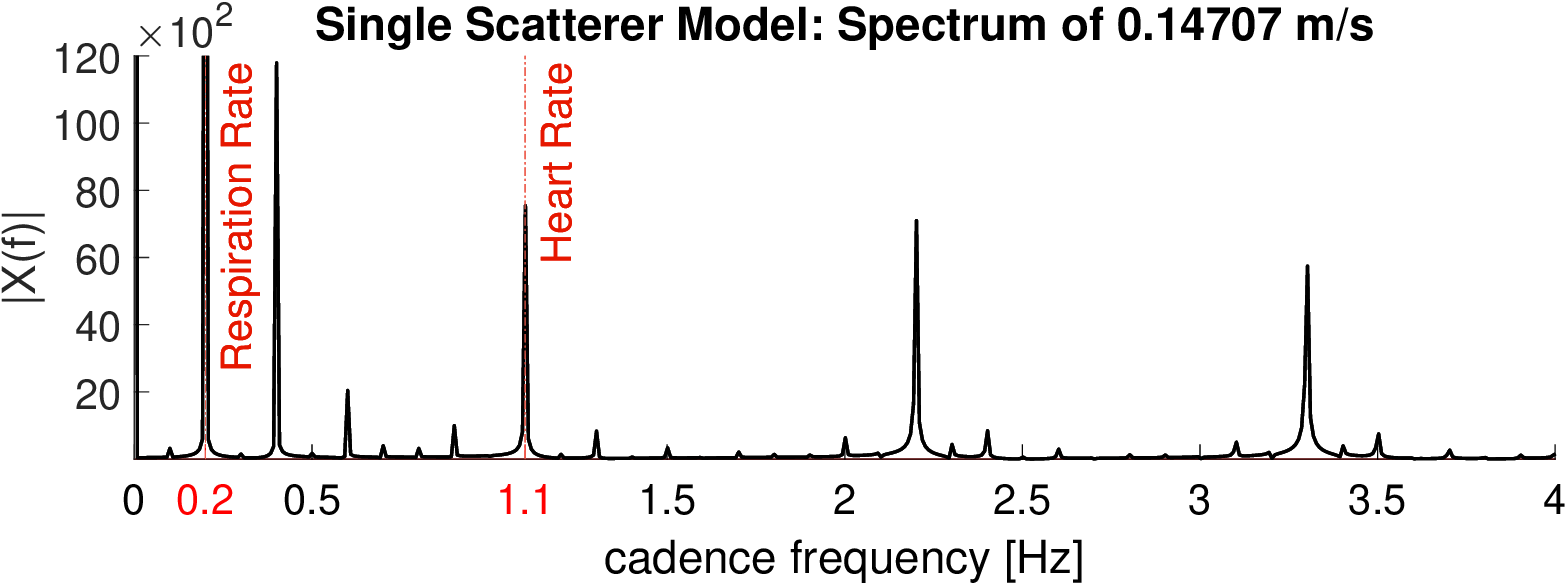}
        \caption{Spectrum of 14.7 cm/s \textit{logarithm} for the Single Point Scatterer}
        \label{fig:single_scatterer_highvel}
    \end{figure}
    \begin{figure}[t]
        \centering
        \includegraphics[width=\columnwidth]{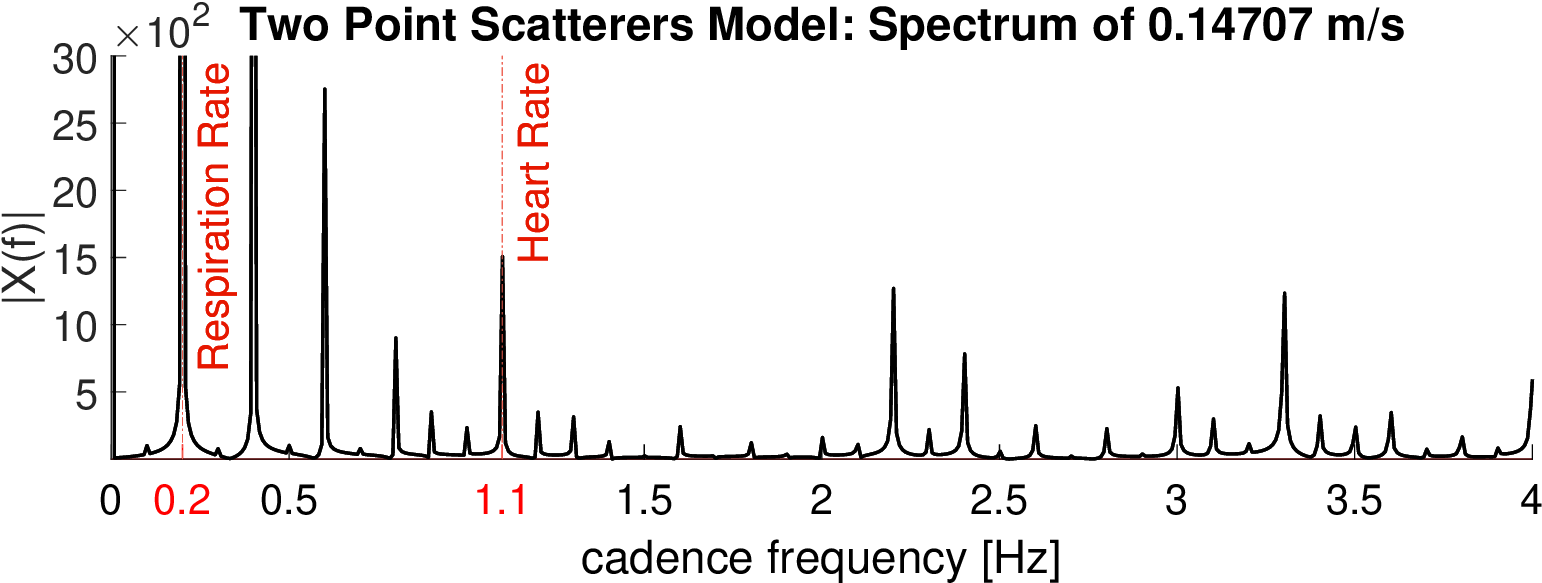}
        \caption{Spectrum of 14.7 cm/s \textit{logarithm} for the Two Point Scatterers}
        \label{fig:two_scatterers_highvel}
    \end{figure}

    \figurename~\ref{fig:single_scatterer_highvel} and \figurename~\ref{fig:two_scatterers_highvel} show the spectrum of the logarithm of~the selected velocity. Apart from the low-frequency peaks at the multiples of 0.2~Hz (due to the respiration), the majority of high peaks are located at multiples of 1.1~Hz, representative of the heart rate.
    
    The analysis confirms that velocity-time maps can be analysed more extensively to yield vital signs. In scenarios where the phase extraction method limits the estimation of the heart rate, the more complex Doppler spectrum provides a better foundation for carrying out the analysis.
    
\section{Conclusion}
\label{sec:conclusion}
    Reported research to date has suggested that the manifestation of the two human vital signs are distributed on the target's surface over different areas and with different effective RCSs. The analysis presented provides evidence that in the goal of estimating both respiration and heart rates simultaneously, it is more appropriate to utilise a two point scatterers model of a living target than the more traditional single point scatterer model. The former treats more general scenarios and this degree of generality is of value in the development of solutions addressing typical real-world applications. The key advantage of the velocity-time map method is that it is capable of extracting both vital signs regardless of which model represents the factual environment.
    
\section*{Acknowledgment}
\label{sec:acknowledgement}
This work was funded by EPSRC under Grant EP/R513349/1.

\balance

\printbibliography

\end{document}